\documentclass[twocolumn,showpacs,amsmath,amssymb,prl]{revtex4}
\usepackage{graphicx}

\begin{document}

%\title{Store heat Information: a Model of Thermal Memory}
\title{Thermal memory: a storage of phononic information}

\author{Lei Wang$^{1,2}$
%\footnote{phywanglei@ruc.edu.cn}
and Baowen Li$^{2,3}$\footnote{Correspondence should be addressed to:
phylibw@nus.edu.sg}} \affiliation{ $^1$ Department of Physics,
Renmin University of
China, Beijing 100872, P. R. China\\
$^2$ Department of Physics and Centre
for Computational Science and Engineering,
 National University of Singapore, Singapore 117542\\
$^3$ NUS Graduate School for Integrative Sciences and Engineering,
Singapore 117597, Republic of Singapore}

\date{25 August 2008}

\begin{abstract}
Memory is an indispensable element for computer besides logic gates.
In this Letter we report a model of thermal memory. We demonstrate
via numerical simulation that thermal (phononic) information stored
in the memory can be retained for a long time without being lost and
more importantly can be read out without being destroyed. The
possibility of experimental realization is also discussed.
\end{abstract}
%original proposed pacs.
%\pacs{07.20.Pe,07.20.-n, 63.22.+m,89.20.Ff, }
%85.90.+h Other topics in electronic and magnetic devices and microelectronics
%89.20.Ff Computer science and technology
%07.20.-n Thermal instruments and apparatus
%07.20.Pe Heat engines; heat pumps; heat pipes
%63.22.+m Phonons or vibrational states in low-dimensional
%structures and nanoscale materials

\maketitle

Unsatiable demands for thermal management/control in our daily life
ranges from thermal isolating to efficient heat dissipation have
driven us to better understand heat conduction from molecular level.
Recent years has witnessed a fast development including theoretical
proposals in functional thermal devices such as thermal diode that
rectifies heat current \cite{diode}, thermal transistor that
switches and modulates heat current \cite{transistor}, heat pump
that carries heat against temperature bias \cite{heatpump}, and
experimental works such as nanotube phonon waveguide
\cite{waveguide}, thermal conductance tuning \cite{thermaltuning}
and nano-scale solid state thermal rectifier
\cite{experimentaldiode}. More importantly, logic operations with
phonons/heat have been demonstrated theoretically\cite{thermalgate},
which, in principle, has opened the door for a brand new subject -
phononics - a science and engineering of processing information with
heat\cite{phononics}. The question arises naturally and immediately:
whether a thermal (phononic) memory that can store information is
possible?

In this Letter, we would like to give a definite answer by
studying the transient process, which in fact
exhibits much richer phenomena than asymptotical stationary state
does, of a non-equilibrium system. This topic has been however rarely studied so far.

Like an electronic memory that records data by maintaining voltage
in a capacitance, a thermal memory stores data by keeping
temperature somewhere. An ideal memory keeps the data forever
without fading. This is never possible for a thermal system because
sooner or later randomness of the heat (fluctuation) will eliminate
any record of the history. However this problem is not serious from
application point of view since we only need to maintain the data
until we refresh it or read it. This is exactly the case for the
widely used Dynamic Random Access Memory (DRAM) that needs {\it
refresh} operation regularly. Any system that keeps temperature
(thus data) somewhere for a very long time might be a candidate for
thermal memory, such as breather and localized harmonic mode {\it
etc}. However perturbation is unavoidable in a thermal system,
especially as the data are read (i.e., the local temperature is
measured). The breather and harmonic mode, even if do not collapse
immediately under the perturbation\cite{breather}, are not able to
recover their original state. Because the energy that is changed by
the perturbation, e.g., the necessary energy exchange in order to
build equilibrium between the system and the thermometer during the
data reading (temperature measuring) process, is not recoverable in
the autonomous systems without an external energy source/sink. We
thus have to turn to a thermal-circuit with power supply, i.e.,
driven by external heat bath.

Like an electric circuit, the temperature and heat current
distributions of a thermal-circuit are determined by Kirchhoff's
laws. In any linear thermal-circuit, i.e. all thermal resistances
are fixed, independent of temperature and/or temperature drop, the
steady state that satisfies Kirchhoff's laws must be unique. This
unique state must be stable, namely under any perturbation the
system will eventually return to this state. However in order to work as
a thermal memory, a thermal-circuit must have more than one stable steady states,
such a bi-stable thermal-circuit is only possible in the presence of
nonlinear thermal devices.

\begin{figure}[ht]
\includegraphics[width=\columnwidth]{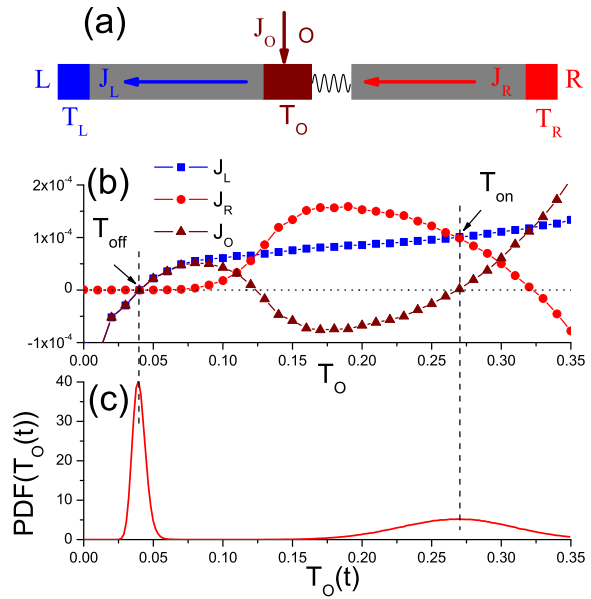}
\vspace{-0.5cm} \caption{\label{fig:JvsTG_Tdis} (Color-online) (a)
Configuration of a thermal memory. Black wavy curve indicates the
weak coupling between particle $O$ and segment $R$, which results in
the {\it negative differential thermal resistance} (NDTR) here. (b)
Heat currents, $J_L$, $J_R$ and $J_O$ versus temperature $T_O$. The
NDTR between particle $O$ and segment $R$ makes $J_R$ increases in a
wide region  when
 $T_O$ is increased, thus makes the curve cross with $J_L$ at
more than one values of temperature $T_O$. (c) PDF of the finite
time temperature $T_O(t)$ of the central particle $O$ in the absence
of control heat bath. Notice the clear correspondence of the two
stable steady states in (b) and the two peaks in (c), which are
indicated by the two vertical dot lines. Since $T_O(t)$ can only
change continuously, the very low PDF in between clearly points out
the low transition probability, thus implies the stability of the
two states ``on'' and ``off''. }
\end{figure}

To make such a thermal-circuit, we consider a one dimensional {\it
nonlinear} lattice that consists of two Frenkel-Kontorova (FK)
segments \cite{FK} sandwich a central particle in between. Its
configuration is shown in Fig.\ref{fig:JvsTG_Tdis}(a) and its
Hamiltonian follows:
\begin{equation}
\label{eq:H} H=H_{L}+H_{R}+H_{int},
\end{equation}
and the Hamiltonian of each segment $L/R$ can be written as
$H_{L/R}$=$\sum_{i=1}^{N_{L/R}}\frac{\dot{x}_{L/R,i}^{2}}{2} +
\frac{k_{L/R}}{2}(x_{L/R,i}-x_{L/R,i-1})^{2}-\frac{V_{L/R}}{(2\pi)^{2}}\cos
2\pi x_{L/R,i}$, with $x_{L/R,i}$ denotes the displacement from
equilibrium position of the $i^{th}$ particle in segment $L/R$. We
have set the mass of each particle to unity. The parameters $k$ and
$V$ are the harmonic spring constant and the on-site potential of
the FK segments. We couple the last particles of segments $L$ and
$R$ to particle $O$ via harmonic springs. Thus
$H_{int}$=$\frac{\dot{x}_{O}^{2}}{2}+\sum_{L/R}
\frac{k_{intL/R}}{2}(x_{L/R,N_{L/R}}-x_{O})^2-\frac{V_{O}}{(2\pi)^{2}}\cos
2\pi x_{O}$. In our numerical simulation, segment $L$ and $R$ each
contains 50 particles. Fixed boundary conditions are applied. Other
parameters are: $K_L$=$K_{intL}$=0.45, $K_R$=0.2, $K_{intR}$=0.05,
$V_L$=10, $V_R$=0, $V_O$=10. The deep valley of on-site potential of
particle $O$, $V_O$, and the weak coupling between segment $R$ and
particle $O$, $K_{intR}$, induce a {\it negative differential
thermal resistance} (NDTR)\cite{transistor} between $O$ and segment $R$,
which is the key for the model of thermal memory. We use a wavy curve to stress this in
Fig.\ref{fig:JvsTG_Tdis}(a). $T_L$ and $T_R$, temperatures of the
power supplies (heat baths) that contacted to the ends of the two
segments, are fixed to $0.04$ and $0.3$, respectively. Without these
energy resource/sink, the system reduces to an autonomous one thus, as has been explained above,
is not able to work as a thermal memory. All heat baths are
simulated by Langevin heat baths \cite{review}.

Because of NDTR, with fixed $T_L/T_R$ and
a control heat bath with adjustable temperature $T_O$ is coupled
to the particle $O$,
there exist more than one values of $T_O$ with which the heat
current from control heat bath to particle $O$, $J_O$, is
zero\cite{transistor}, as shown in Fig.\ref{fig:JvsTG_Tdis}(b).
Suppose the temperature of $O$ is initially set to either
``$T_{on}$'' or ``$T_{off}$'' by the control heat bath. After the
control heat bath is removed the state will remain unchanged for a
long time, in spite of the thermal fluctuation. Because as the consequence
of the small perturbation which slightly changes the temperature $T_O$,
the changes of heat currents $\delta J_L$ and $\delta J_R$ always pull $T_O$ back,
thus these two states are ``stable''. It is
easy to check the other steady state in between is however unstable.

In the equilibrium state of an Hamiltonian system with ergodicity,
temperature can be defined as twice of the average kinetic energy
per degree of freedom. In order to study transient process, the
``finite time temperature'' is naturally defined as, for a 1D
system, twice of the average kinetic energy in a time window $\delta
t$, i.e.,: $T(t)\equiv \frac1{\delta t}\int_{t-\frac{\delta
t}2}^{t+\frac{\delta t}2} v^2(t') dt'.$ Hereafter, any temperature
which is a function of time $t$ means the finite time temperature.
Due to the thermal fluctuation, even in a steady state, finite time
temperature has different value at different time window. The
probability density function (PDF) of the finite time temperature is
thus defined accordingly. If the time window $\delta t$ is chosen to
be too long, the PDF approaches a $\delta$ function locates at
``infinite time'' temperature $T$, while if $\delta t$ is too short,
big thermal fluctuation dominates thus the distribution is more or
less a Maxwell energy distribution. In the medial cases PDF presents much
more information other than the value of temperature $T$ itself. In
this Letter, $\delta t$ is fixed to $10^4$ dimensionless time units.
Since the frequency of the system we studied is about 0.2
\cite{transistor}, this period of time covers about 2,000
oscillations, thus is a microscopical long time.

Since ``on'' and ``off'' are two stable states, the finite
time temperature of the particle $O$ in the absence of control heat
bath is expected to stay around $T_{on}$ and $T_{off}$ with much higher probability than elsewhere.
In Fig.\ref{fig:JvsTG_Tdis}(c) the locations of the two peaks clearly confirms this expectation.
By adjusting parameters of the system, the values of $T_{on}$ and $T_{off}$
and the peaks shift simultaneously but always coincide with each other.

\begin{figure}[ht]
\includegraphics[width=\columnwidth]{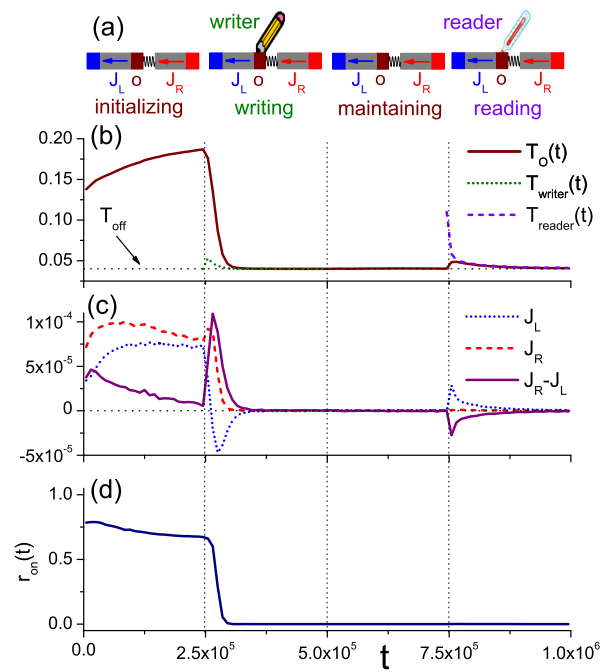}
\vspace{-0.8cm} \caption{\label{fig:w_r_off} (Color-online) {\bf A
writing-reading process for an ``off'' state}. Vertical dot lines
separate different stages to guide eyes.(a) Illustration of the four
stages. (b) Temperatures of the central particle $T_O(t)$, writer
$T_{writer}(t)$ and reader $T_{reader}(t)$. (c) Heat currents
through the two segments, $J_{L}$ and $J_{R}$.
 It can be seen that at the beginning of the reading stage,
 the net current $J_{R}$-$J_{L}$ changes to negative,
 namely, energy is absorbed from $O$ thus $T_O(t)$ is cooled down back to the low temperature $T_{off}$.
(d) After the writing stage $r_{on}(t)$ keeps very close to zero,
 namely nearly all samples stay at the ``off'' state,
 even under the not-very-small perturbation from the hot reader. }
\end{figure}

We have so far confirmed the stability of the two states ``on'' and
``off'' of the model. In the following, we shall demonstrate a
complete, four-stage, writing-reading process of the thermal memory,
as is shown in Fig.\ref{fig:w_r_off}(a). During the whole process
the left and right ends of the memory are always connected to two
power supply heat baths with temperatures $T_L$$=$$0.04$ and
$T_R$$=$$0.3$, respectively. {\bf Stage 1: initializing the memory}.
Each particle is initially set a velocity chosen from a Gaussian
distribution corresponding to a temperature in $(T_L, T_R)$.  After
some time the system approaches to an asymptotic state, thus
$T_O(t)$ saturates, see Fig.\ref{fig:w_r_off}(b). {\bf Stage 2:
writing the datum into the memory}.  In our simulation, the
``writer'' is simulated by a FK lattice with $N$=10 particles and
identical parameters with segment $L$. It is connected via a linear
spring with $k$=1.0, at its one end, to the particle $O$ of the
memory. The temperature of the writer is initially set to $T_{off}$
and the other end is connected to a heat bath with temperature
$T_{off}$ also. In a short time the writer cools down $T_O(t)$ to
$T_{off}$, namely the datum ``off'' has been written into the
memory. {\bf Stage 3: maintaining the datum in the memory}. In this
stage, the writer is removed from the memory. And we can see from
Fig.\ref{fig:w_r_off}(b) that $T_O(t)$ remains nearly unchanged
during this stage, which means that the datum stored in the memory
can last for a long time. {\bf Stage 4: reading out the datum from
memory}. A ``reader''(thermometer) is used to read the datum out
from the memory. Here the ``reader'' is simulated by the same FK
lattice as the writer. It is connected, to the particle $O$ of the
memory. The temperature of the reader is initially set to a middle
temperature between $T_{on}$ and $T_{off}$ (in this Letter it is
0.11). Different from the writer, the reader is not connected to any
heat bath. This stage is the most important and most interesting
one. We see that at the beginning the particle $O$ is temporarily
warmed up because the reader is hotter. However, since the ``off''
state is stable, $T_O(t)$ and also the temperature of the reader are
cooled down to $T_{off}$ shortly, implying that the datum has been
successfully read out. To gain a clearer picture, we show the heat
currents through the two segments $J_{L}$ and $J_{R}$ in the
Fig.\ref{fig:w_r_off}(c). At the beginning of the reading stage
since the temperature of particle $O$ is warmed up by the reader, as
a response of the increase of $T_O(t)$, $J_L$ increases a lot
whereas $J_R$ increases only very little, thus the net current
$J_R$-$J_L$ changes from zero to negative, i.e., the power supply
(left end heat bath) absorbs heat from the particle $O$. This is the
reason that $T_O$ can be cooled down and recovered to $T_{off}$
automatically.

The results shown in Fig.\ref{fig:w_r_off} are obtained from an
ensemble average over $20,000$ independent samples. For a model of
thermal memory, the ratio of samples that keep the correct data is
an important feature. To describe this quantitatively, we calculate
the ratio of ``on'' state: $r_{on}(t)\equiv N_{on}(t)/N$, while
$N_{on}(t)$ is the total number of samples stay at ``on'' state at
time $t$ and $N$ is the total number of samples. We define, at time
$t$, a sample is at ``on'' state if the finite time temperature of
its particle $O$ is greater than a critical value $T_c$=0.11. we
show $r_{on}(t)$ during the whole process in
Fig.\ref{fig:w_r_off}(d). It is clearly seen that after the writing
stage, $r_{on}(t)$ always keeps at a very low level, even under the
not-very-small perturbation from the hot reader. The discrepancy
between $r_{on}(t)$ and 0 is in the order of $10^{-4}$, which means
two or three ``errors'' among the 20,000 samples, is even
indistinguishable in the figure by eye, namely the memories
successfully maintain the data ``off'' that are initially set by the
writer, the data are hardly destroyed even after being read!

\begin{figure}[ht]
\includegraphics[width=\columnwidth]{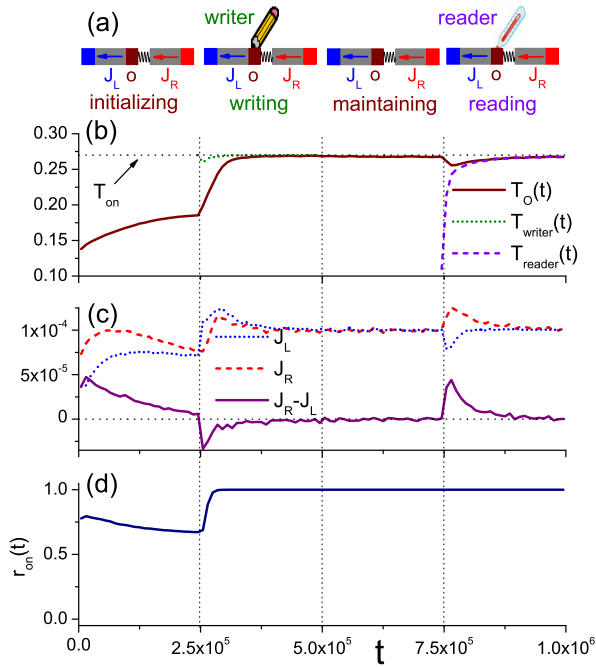}
\vspace{-0.8cm} \caption{\label{fig:w_r_on} (Color-online) {\bf A
writing-reading process for an ``on" state}.  (a) Illustration of
the four stages. (b) Temperatures of the central particle $T_O(t)$,
writer $T_{writer}(t)$ and reader $T_{reader}(t)$. (c) Heat currents
 through the two segments, $J_{L}$ and $J_{R}$. At the beginning of
the reading stage, the net current $J_{R}$-$J_{L}$ is positive,
namely provides energy to $O$ thus warms it up back to the high
level $T_{on}$. (d) After the writing stage $r_{on}(t)$ remains very
close to one,  namely, nearly all samples stay at the ``on'' state,
even under the not-very-small perturbation from the cold reader.}
\end{figure}

It has been pointed out that, to be a memory, the system
must have more than one stable steady states around which
the above writing-reading process can be completed, thus
in the following we demonstrate the writing-reading process
around the other stable steady state, the ``on'' state
in Fig.\ref{fig:w_r_on}. The
difference from Fig.\ref{fig:w_r_off} is that in the second stage
the writer is set to temperature $T_{on}$ and connected to a heat
bath with temperature $T_{on}$. We see in this stage $T_O(t)$
changes to $T_{on}$ shortly. In the consequent maintaining stage $T_O(t)$
keeps unchanged at $T_{on}$. In the reading stage: the reader is
initially set to the same temperature as before. We see at the
beginning because this time the reader is colder, the particle $O$
is temporarily cooled down, as a response $J_{L}$ decreases while $J_{R}$
increases, thus the net current $J_{R}$-$J_{L}$ becomes positive,
which warms up the particle $O$ (also the reader) and recovers its
temperature to $T_{on}$. The datum ``on'' is again read out
successfully. As for $r_{on}(t)$, see Fig.\ref{fig:w_r_on}(d). This
time $r_{on}(t)$ keeps at a very high level. The discrepancy between
$r_{on}(t)$ and 1 is again at the order of $10^{-4}$, namely the
datum ``on'' can be successfully maintained in and read out from
memory.

Finally we would like to discuss the lifetime of data in the thermal
memory. In an electric DRAM, since real capacitors leak charge, the
data will eventually fade unless the capacitor charge is refreshed
periodically (about $\sim 100/s$). Similarly, we find a few
``errors'' among the many samples that we studied at the end of the
writing-reading process. Suppose the data fading process is a
Poisson process then the average lifetime of the data is roughly
$5*10^9$ dimensionless time units which contains about $10^9$
periods of oscillation. This corresponds to 100 $\mu s$ if the
memory is made of carbon nanostructures such as
nanotube\cite{carbon}, in which the key of the model, NDTR has been
found numerically\cite{WuLi0708}. This lifetime is unsatisfying when
compared with an electronic DRAM. However this value can be easily
and greatly enlarged by parallel combining more identical memories
together. The dynamics of finite time temperature of particle $O$,
$T_O(t)$ around either of the two stable steady states can be
roughly described by an autoregressive diffusion process, say, the
simplest one, Ornstein-Uhlenbeck (OU) process: $\dot X=-\lambda
X+\sigma \xi(t)$. Where $X$ corresponds to $T_O(t)$ relative to
$T_{on}$ or $T_{off}$. $\lambda$$>$$0$ represents the effect of the
response net heat current that pulls $X$ back to zero and $\sigma
\xi(t)$ is a Gaussian white noise with zero mean and fixed variance
$\sigma^2$ which describes fluctuation of the heat current. The mean
first passage time (MFPT) through a certain boundary $X=\Delta X$
(which denotes $T_c-T_{off}$ or $T_{on}-T_c$) conditional upon
$X(0)=0$ represents the average lifetime of the data in the memory.
In the limiting case that $\frac{\sqrt{\lambda}\Delta
X}{\sigma}$$>$$>$$1$,
MFPT$\propto$$\frac{1}{\lambda}\mbox{erfi}(\frac{\sqrt{\lambda}
\Delta X}{\sigma})$, where $\mbox{erfi}(x)$$=$$\int_0^x e^{x'^2}
dx'$ \cite{MFPT_OU}. MFPT diverges rapidly as $\sigma^2$ decreases.
Parallel combining $n$ identical memories together decreases
$\sigma^2$ by $n$ times while keeps $\lambda$ unchanged, thus
increases the average lifetime of data greatly. Such a fast
divergence tendency of MFPT has also been widely found in a class of
diffusion processes with steady state distribution \cite{OUexample},
thus the above conclusion is not limited to the specified process,
the OU process.

In summary, we have demonstrated the feasibility of thermal memory
from a bistable thermal-circuit. The information stored in such a
thermal memory can last very long time and, more importantly it is
self-recoverable under the not-very-small perturbation from the
thermometer when the data are read. With the rapid developing
nano-technology \cite{waveguide,thermaltuning,experimentaldiode},
the thermal memory should be realized in nanoscale systems
experimentally in a foreseeable future.

The work is supported in part by the start-up grant of Renmin
University of China (LW) and Grant R-144-000-203-112 from Ministry
of Education of Republic of Singapore, and Grant R-144-000-222-646
from NUS.

\end{document}